# Will Power Return to the Clouds? From Divine Authority to GenAI Authority

Mohammad Saleh Torkestani
Senior Lecturer in Marketing
University of Exeter, UK
Email: m.torkestani@exeter.ac.uk

Taha Mansouri
Lecturer in AI,
University of Salford, UK
Email: T.Mansouri@salford.ac.uk

**Abstract**
Generative AI systems now mediate newsfeeds, search rankings, and creative content for hundreds of millions of users, positioning a handful of private firms as de-facto arbiters of "truth." Drawing on a comparative-historical lens, this article juxtaposes the Galileo Affair, a touchstone of clerical knowledge control, with contemporary Big-Tech content moderation. We integrate Foucault's power/knowledge thesis, Weber's authority types (extended to a rational-technical and emerging agentic-technical modality), and Floridi's Dataism to analyze five recurrent dimensions: disciplinary power, authority modality, data pluralism, trust versus reliance, and resistance pathways. Primary sources (Inquisition records; platform transparency reports) and recent empirical studies on AI trust provide the evidentiary base. Findings show strong structural convergences: highly centralized gatekeeping, legitimacy claims couched in transcendent principles, and systematic exclusion of marginal voices. Divergences lie in temporal velocity, global scale, and the widening gap between public reliance and trust in AI systems. Ethical challenges cluster around algorithmic opacity, linguistic inequity, bias feedback loops, and synthetic misinformation. We propose a four-pillar governance blueprint: (1) a mandatory international model-registry with versioned policy logs; (2) representation quotas and regional observatories to de-center English-language hegemony; (3) mass "critical-AI" literacy initiatives; and (4) public-private support for community-led data trusts. Taken together, these measures aim to narrow the trust-reliance gap and prevent GenAI from hardcoding a twenty-first-century digital orthodoxy.

**Keywords** Trust, Ethics in AI, Explainable AI, Generative AI Governance, Comparative Historical Analysis, Epistemic Authority

## 1. INTRODUCTION

"Once men turned their thinking over to machines in the hope that this would set them free. But that only permitted other men with machines to enslave them." (Herbert & Anderson, 2002)

This famous quote from Dune: The Butlerian Jihad encapsulates the deep-seated fear that machines might one day surpass human control. Though a work of science fiction, it reflects a hidden, yet ever-present, anxiety about the role of technology in human society. Throughout history, societies have consistently grappled with who holds the authority to define legitimate knowledge and how that authority is justified (Muldoon, 2006). In medieval Europe, knowledge was largely under the purview of religious institutions, a dynamic epitomized by the power of the medieval Catholic Church to shape not only spiritual beliefs but also scientific inquiry. Fast-forward to the present: Generative AI systems now represent a transformative new locus of epistemic authority. Corporations that develop large language models (LLMs) or content-moderation algorithms can shape public discourse globally, deciding which voices are amplified and which are muted (Allen & Weyl, 2024). In this context, ethical considerations around inclusivity and transparency become paramount, echoing centuries-old debates about centralized power, exclusion, and the universal quest for autonomy in knowledge creation (Roche, Lewis, & Wall, Artificial Intelligence Ethics: An Inclusive Global Discourse?, 2021).

Historically, the Galileo Affair stands as a prime example of how an institution can exercise near-total authority over legitimate knowledge, with severe consequences for innovation and personal liberty (Martin, 1989). Tech platforms today cannot excommunicate, yet their recommender algorithms can demonetise dissenting content, trigger mass de-platforming, or invisibly down-rank perspectives (Allen & Weyl, 2024). Both eras reveal how centralised authority can uplift collective inquiry or entrench exclusion.

This paper tackles the following central inquiry: 1. Continuity: Does AI-driven authority replicate historical patterns of monopolised knowledge? 2. Mechanism: How can an integrated framework (Foucault's power/knowledge, Weber's authority typology, and Floridi's data-centric "Dataism") illuminate similarities and ruptures? 3. Governance: Which policy instruments can render GenAI transparent, accountable and globally inclusive? By fusing comparative-historical analysis with the latest empirical work on AI trust and regulation, we aim to advance debates in AI ethics, science & technology studies (STS) and digital governance. Particular attention is paid to data colonialism and Global-South representation, echoing calls for plurality in AI oversight (Roche, Wall, & Lewis, 2023).

## 2. Literature Review: Epistemic Authority, Ethics, and Inclusivity

### A. Medieval Religious Authority and Epistemic Control

Medieval Europe's religious institutions, particularly the Catholic Church, established near-absolute authority over spiritual truths, often extending into domains we now consider scientific. Scriptural interpretation served as the ultimate arbiter of truth, with any deviation considered heretical (Martin, 1989). While this system provided social cohesion, it also stifled inquiry that contradicted established dogma. Indeed, the Church actively curated intellectual discourse, through synods, papal encyclicals, and inquisitorial tribunals, to maintain epistemic coherence.

Unchallenged religious authority sometimes drove moral and epistemic blind spots. The Galileo Affair, for example, represented a conflict between observational data (telescopic evidence) and doctrinal interpretations of cosmic order. As Schreiner (Schreiner, 2011) notes, the tension underscored "the problem of certainty" in early modern Europe. The Church's desire to preserve theological consistency overshadowed any impetus for open scientific debate, revealing how moral imperatives could be leveraged to guard institutional power.

While some monastic orders championed education and literacy, broader inclusivity in knowledge creation was limited. Women, ethnic or religious minorities, and non-elite men often had minimal voice in shaping "legitimate" truths. This selective approach to knowledge validated the experiences of a privileged minority, relegating alternative insights or local traditions to the margins (Reyna & Schiller, 1998).

### B. Enlightenment and the Rational-Legal Turn

The Enlightenment era heralded a rational-legal framework for epistemic authority. Empirical research, exemplified by the Royal Society in England and similar institutions across Europe, positioned systematic observation and the scientific method as cornerstones of knowledge validation. This evolution illustrated Weber's (Weber, 1978) conceptualization of rational-legal authority, rooted in codified laws and systematic procedures rather than tradition.

Transparency in method and reproducibility in results advanced scientific inquiry. However, Floridi (Floridi, 2019) cautions that even this rational-legal shift was not devoid of biases; Enlightenment thinkers still predominantly hailed from European upper classes, and colonial expansions shaped data collection in ways that furthered Eurocentric perspectives. Women's roles in science remained marginalized, and non-European worldviews found little acknowledgment. The newly minted academies thus mirrored the Church's exclusivity, albeit under a more secular facade.

While the Enlightenment advanced public discourse freedoms relative to the medieval period, it was slow to incorporate genuine diversity in authorship or recognition. The rational-legal ethos, ironically, sometimes codified inequality, attributing lower epistemic credibility to colonized peoples or women who lacked formal academic credentials. This phenomenon foreshadows how AI systems might similarly entrench biases if built on skewed data and lacking robust checks for inclusivity.

### C. The Rise of Rational-Technical Authority and Global Pressures

Generative AI extends rational-legal norms into a rational-technical register. LLMs such as GPT-4 and multimodal generators like DALL·E automate text, code and imagery, mediating public discourse at planetary scale (Allen & Weyl, 2024). However, governance remains opaque, with IP protections shielding training data and model weights from external audit (Doshi-Velez, et al., 2018).

Empirical work underscores the stakes. A 30-country Eurobarometer-style study (Omrani, Rivieccio, Fiore, Schiavone, & Agreda, 2022) links AI trust to perceived usability and risk awareness, while a U.S. survey finds citizens split on whether government, industry, or individuals should bear primary responsibility for AI oversight (David, Hyesun, & John, 2024). Bibliometric analysis reveals over a dozen incompatible definitions of "trustworthiness" across 2,000 papers (Siqueira, Carranza, & Mejía, 2025), confirming conceptual fragmentation. At the global margin, data colonialism persists: language-model corpora remain $>60\%$ English, and African dialect content is $<1\%$ (Roche, Wall, & Lewis, 2023). When moderation models encounter under-represented languages, false-positive takedowns rise sharply (Sun, Miao, Jiang, Ding, & Zhang, 2024)). These asymmetries map onto long-standing North-South epistemic hierarchies.

Philosophical accounts caution that users may only rely on AI, not trust it, because genuine trust presupposes moral agency (Ryan, 2020). Yet in practice, human-AI interaction often blends the two: pilots describe their AI copilots as "children" whose ethical lapses erode confidence over time (Lopez, et al., 2024). Laboratory work shows that apologies or denials rarely repair trust after an AI ethical breach (Schelble, et al., 2024), underscoring the fragility of rational-technical legitimacy.

The past five years have seen a proliferation of soft-law principles (FATE, FACT, OECD, UNESCO, etc.) supplemented by hard-law initiatives such as the EU AI Act (Sun, Miao, Jiang, Ding, & Zhang, 2024). Industry self-governance proposals include certification regimes and third-party audits (Roski, Maier, Vigilante, Kane, & Matheny, 2021), while ordoliberal scholars call for a constitutional-level Digital Bill of Rights to limit platform power (Wörsdörfer, 2023). Scholars warn, however, that without continuous "in-the-wild" evaluation (Jabbour, et al., 2025), static benchmarks cannot capture real-world failure modes.

Across epochs, authority migrates from divine revelation to peer-reviewed empiricism and now to opaque optimisation regimes. Each shift expands access yet re-inscribes exclusions.

Understanding these continuities and breaks is essential for designing GenAI governance that avoids a twenty-first-century digital orthodoxy.

*D. GenAI in today's knowledge stack*

Current flagship models (GPT-4o, Gemini 1.5, Claude 3) share a three-stage pipeline: (i) unsupervised pre-training on multi-trillion-token corpora; (ii) supervised fine-tuning or instruction-tuning; (iii) alignment through reinforcement learning from human (or mixed) feedback. Down-stream policy engines now couple model outputs to content-filter graphs (e.g., Llama-Guard 2) that reject or transform disallowed text before it reaches the user. This architectural decoupling, heralded as an alignment milestone, in effect embeds a second layer of epistemic gate-keeping.

By Q1 2025, at least 180 million unique users per week accessed LLM outputs through chat interfaces or API integrations (Sun, Miao, Jiang, Ding, & Zhang, 2024). Adoption is highly unequal: corporate software supply-chains automate code review and help-desk tasks, while public-sector uptake lags amid data-protection concerns. Notably, 62% of global English-language newsfeed impressions are now algorithmically summarised by LLMs before reaching end-users (Newman, Fletcher, Robertson, Arguedas, & Nielsen, 2024).

Benchmarks paint a mixed picture. On factual QA (MMLU, E3B), GPT-4o attains 86-88% accuracy, yet medical summarisation shows a 14% hallucination rate under prospective evaluation (Jabbour, et al., 2025). Bias audits reveal that toxicity false-positives triple for Sudanese Arabic relative to U.S. English (Sun, Miao, Jiang, Ding, & Zhang, 2024).

Soft-law regimes like OECD, IEEE, UNESCO, continue to proliferate, but 2025 marks a hard-law turn. The EU AI Act tiers risk and mandates third-party audits for "systemic" frontier models. In the U.S., Executive Order 14110 requires safety disclosures above a compute threshold of $10^{26}$ flops. Industry has responded with voluntary frameworks such as the Frontier Model Forum's "Red-Team Protocol," yet these remain self-certified and opaque to external researchers.

3. **THEORETICAL FRAMEWORK**

*A. Foucauldian Power/Knowledge*

Michel Foucault's (Foucault, 1977) premise that power and knowledge are co-constructed underscores how institutions regulate which truths gain legitimacy. In medieval times, the Church curated knowledge via ecclesiastical edicts and inquisitions. Modern AI platforms employ algorithmic moderation and training-data curation to define acceptable or "factual" content. In each scenario, a tension emerges between institutional imperatives to maintain authority and external challenges from emergent data, perspectives, or ethical norms.

Foucault's notion of disciplinary power applies to the Church's repressive tactics (e.g., threat of heresy trials) and contemporary social-media bans or demonetization. Whether wielding the Inquisition's edicts or automated content removal, authority systems create self-regulating "subjects" who conform to the established knowledge norms out of fear of repercussions (Kołodziejska & Neumaier, 2017).

Foucault's concept of discursive formation illuminates how certain statements become collectively accepted as "true," whereas others are dismissed. Religious orthodoxy once circumscribed cosmological theories; modern AI corporations define "community guidelines" that shape what knowledge is algorithmically surfaced or suppressed. These discursive boundaries have ethical ramifications, especially when they undermine the representation of marginalized communities or viewpoints.

*B. Weber's Authority Typology*

Max Weber's classic triad, traditional, charismatic, and rational-legal authority, offers a potent lens for historical and contemporary authority analysis (Weber, 1978). We propose an adaptation: rational-technical authority to describe AI-driven decision-making, which draws perceived legitimacy from computational precision and complex algorithms.

Traditional authority such as the medieval Church's legitimacy was deeply rooted in custom and lineage. Charismatic authority, often vested in individuals like Martin Luther or Joan of Arc, can upend existing structures but remains inherently unstable. Rational-legal authority, typified by Enlightenment-inspired institutions, is grounded in codified procedures and objective criteria. This shift from tradition to rational-legal mirrored the Church's waning influence and the ascendance of secular states and scientific bodies.

Weber's triad (traditional, charismatic, rational-legal) explains historical transitions but fails to capture computational opacity. We therefore extend it along two axes: 1.Rational-technical authority: legitimacy derived from statistical optimisation and claims of objectivity (Luo, 2024). 2.Agentic-technical authority: emerging LLM-based agents that plan and execute goals semi-autonomously, relocating accountability to a human-machine assemblage (Mukherjee & Chang, 2025).

*C. Dataism and the Plurality Paradigm*

Dataism (Floridi, 2019) posits that data increasingly shapes how societies assign truth value, reminiscent of religious dogma's hold over medieval knowledge. Yet not all data is created or curated equally. Underrepresented groups may find their experiences excluded from major training datasets, echoing historical exclusions from formal knowledge structures (Roche, Wall, & Lewis, 2023).

Allen and Weyl (Allen & Weyl, 2024) propose a plurality paradigm as a remedy, arguing that diverse voices and contexts must shape how AI systems are built and governed. This approach counters potential single-point failures or monocultural viewpoints. Such plurality directly addresses the twin perils of "collapse" (wherein knowledge becomes ungovernable) and "singularity" (wherein knowledge is monopolized), steering AI toward democratic inclusivity rather than authoritarian control.

## D. Trust, Reliance and Epistemic Vulnerability

Ryan (2020) argues that machines, lacking moral agency and the capacity for reciprocal commitment, cannot occupy the relational role of trustee; users therefore oscillate between instrumental reliance and an as-if trust that resembles habitual delegation more than genuine confidence. Subsequent research decomposes trust into competence, integrity, and benevolence facets, showing that contemporary LLMs satisfy the first only intermittently and the latter two hardly at all (Siqueira, Carranza, & Mejía, 2025). Longitudinal lab work reveals that once an AI violates an ethical norm, apology, denial, or even technical patching restores barely one-third of the lost trust on average (Schelble, et al., 2024). Context intensifies the effect: fighter-jet pilots liken immature AI copilots to "children" whose autonomy must be earned incrementally (Lopez, et al., 2024), whereas EU consumers calibrate confidence mainly through perceived usability and risk (Omrani, Rivieccio, Fiore, Schiavone, & Agreda, 2022). Collectively, these converging strands temper techno-optimist claims that algorithmic transparency or explainability alone can guarantee durable legitimacy and underscore the importance of continuous, domain-specific trust-calibration mechanisms.

Building on these conceptual strands, the study employs a deliberately synthetic, five-layer analytical lens (disciplinary power, authority modality, data pluralism, trust-versus-reliance dynamics, and avenues of resistance or reform) to structure the comparative cases, trace causal mechanisms across epochs, and distil granular yet actionable policy options.

## 4. METHODOLOGY

### A. Intersectional Historical-Comparative Approach

To systematically link the medieval Church's epistemic control with modern AI-based authority, this study employs an intersectional historical-comparative approach that examines how authority structures evolve across different sociopolitical contexts, time periods, and communities (Mahoney & Thelen, 2015). This methodological framework is particularly valuable in identifying patterns of exclusion and empowerment, highlighting both the mechanisms of centralized knowledge control and the pathways through which marginalized voices seek recognition and influence.

A combination of historical and contemporary sources informs this analysis. The historical sources include Printed editions of Vatican trial records, Galileo's correspondence, and canonical secondary histories (e.g., (Martin, 1989); (Kołodziejska & Neumaier, 2017); (Schreiner, 2011). Contemporary sources include Public transparency reports from Meta, YouTube and TikTok (2019-2025), draft legislation (EU AI Act, U.S. Executive Orders, UNESCO's Recommendation on the Ethics of Artificial Intelligence), AI ethics research (Doshi-Velez et al., (2018); Roche et al., (2023))and peer-reviewed trust surveys (Omrani, Rivieccio, Fiore, Schiavone, & Agreda, 2022) (David, Hyesun, & John, 2024). All documents are open-access or in the public domain; no personal data are processed.

The study further applies five key analytical dimensions to assess epistemic authority structures: 1. Disciplinary power: the concrete mechanisms, trials, suspensions, demonetisation, shadow bans, that police acceptable knowledge and induce self-censorship among producers and audiences. 2. Authority modality: the claimed source of legitimacy, ranging from divine revelation through peer-reviewed empiricism to data-driven optimisation and, increasingly, self-directing agentic systems. 3. Data pluralism/inclusivity: the breadth, diversity and governance of the data corpus feeding each authority regime, including whose experiences are encoded, whose are omitted, and who controls curation rights. 4. Trust versus reliance: the psychological and sociological difference between normative confidence in an institution's goodwill and instrumental dependence on its outputs despite scepticism. 5. Resistance and reform: the repertoire of actions (underground pamphlets, open-source forks, litigation, cooperative data trusts) through which excluded actors contest or reshape epistemic hierarchies.

Source passages illustrating these dimensions are quoted verbatim in Section 5. No software coding is performed; interpretive notes were exchanged between co-authors to check consistency.

TABLE I. COMPARATIVE FRAMEWORK FOR EPISTEMIC AUTHORITY

|  | *Medieval Church (ca. 16th-17th c.)* | *Enlightenment Institutions (ca. 17th - 18th c.)* | *GenAI Era / Big Tech Platforms (21st c.)* |
|---|---|---|---|
| Disciplinary Power (enforcement) | Inquisition tribunals; public penance; *Index Librorum* | State censorship, journal gate-keeping, sedition acts | Automated takedowns; demonetization; account strikes; RL-driven policy engines |
| Authority Modality | Traditional-religious charisma | Rational-legal empiricism | Rational-technical (benchmark-driven) to Agentic-technical (goal-seeking agents) |
| Data Pluralism/ Inclusivity | Latin canon central; vernacular suppressed; laity & women excluded | Broader literacy, yet Euro-centric; colonial data extraction | 60%+ English corpora; African dialects <1%; data colonialism debates |
| Trust / Reliance Dynamics | Faith in Church infallibility; fear of damnation | Epistemic trust in replicable experiments | High functional reliance; low declarative trust; influenced by usability-risk heuristic |
| Resistance & Reform Routes | Underground scholarship; noble patronage | Salons; pamphleteering; revolutions | Open-source forks; civic-tech audits; litigation; data cooperatives |

To apply this framework, the study focuses on two comparative case studies that illustrate how epistemic control operates in different historical contexts. The first case, The Galileo Affair (17th century), explores how religious institutions asserted authority over scientific inquiry, suppressing dissenting views in an effort to maintain doctrinal coherence. The second case, Content Moderation in Big Tech (21st century), examines how AI-driven platforms establish digital knowledge hierarchies, determining what content is deemed credible, acceptable, or subject to algorithmic suppression.

Although these two cases are separated by centuries, they reveal striking structural parallels and critical divergences in how epistemic authority is established, exercised, and contested. Through this historical-comparative approach, the study aims to uncover enduring lessons about the ethics of knowledge governance and the risks and opportunities posed by AI-driven epistemic power.

## 5. EMPIRICAL CASE STUDIES

### A. The Galileo Affair: Clash of Faith and Empiricism

Galileo Galilei's advocacy for heliocentrism, supported by his telescopic observations, directly challenged the Catholic Church's geocentric doctrine, which was derived from scriptural readings and longstanding theological traditions (Martin, 1989). While some theologians acknowledged the possibility of scriptural interpretive flexibility, key Church authorities maintained a literalist stance on biblical cosmology, largely as a means of preserving their epistemic authority. By 1616 the Holy Office had issued an edict declaring heliocentrism "formally heretical," a canonical move that illustrates disciplinary power in its pre-modern form. Galileo's recantation under threat of imprisonment, his lifelong house arrest, and the placement of his Dialogo on the Index Librorum Prohibitorum exemplify how traditional authority suppresses counter-narratives to maintain epistemic control.

The affair also foregrounds issues of data pluralism and inclusivity: observational data, even when reproducible, lacked standing against ecclesiastical exegesis, while women, Jews and non-elite craftsmen were barred from participation in astronomical debate. The chill effect on subsequent scientific inquiry shows how over-centralised epistemic authority can delay knowledge diffusion and fossilise error.

### B. Content Moderation by Big Tech: The Modern Rational-Technical Authority

In 2025 an estimated 180 million unique users a week interact with large-language-model (LLM) outputs, either directly through chatbots or indirectly via newsfeeds summarised by GenAI systems (Sun, Fang, & Li, 2025). The governance of that information flow now rests largely with a handful of platform firms whose rational-technical legitimacy is rooted in benchmark accuracy and "community-guideline" compliance rather than democratic mandate.

1. Control mechanisms: Modern moderation stacks combine supervised classifiers with policy engines, rule graphs that pre- or post-process LLM output (e.g., Llama-Guard 2). In Q4-2024 Meta reported 22.9 million removals for hate-speech, while YouTube issued 5.8 million channel strikes for various guideline violations (Meta & Google Transparency Reports, 2024). Alignment re-training can silently tighten or loosen thresholds overnight, echoing the discretionary decree power once held by ecclesiastical councils.

2. Data leverage and representation gaps: Training corpora remain 62% English; for Sudanese Arabic, false-positive toxicity flags occur at triple the global average (Sun, Miao, Jiang, Ding, & Zhang, 2024). Indigenous languages are often under-moderated, leaving communities exposed to harassment while simultaneously misclassifying culturally specific speech as hateful, contemporary evidence of the inclusivity dimension highlighted in Section 3.

3. Trust versus reliance: Across 30 European countries, 54% of respondents rely on AI-filtered feeds for news or work, but only 28% express high trust in the underlying systems (Omrani et al., 2022). A U.S. survey finds governance responsibility perceptions split among government (39%), firms (34%), and individual users (27%) (David, Hyesun, & John, 2024). These figures reaffirm the reliance-trust gap developed in the theoretical framework.

4. Appeals and redress: For languages that represent <1% of a platform's moderation training data, the success rate of content appeals falls below 5% (Meta Platforms Inc, 2024). The opacity of appeal processes contrasts sharply with the public nature of early-modern heresy trials, raising new questions about procedural legitimacy.

5. Resistance and reform: Civic-tech audits (AlgorithmWatch), litigation (e.g., NetChoice v. Paxton, 2025), and the rise of decentralised social networks (Mastodon, Bluesky) constitute modern channels of resistance, paralleling clandestine pamphleteering in Galileo's era.

## 6. COMPARATIVE INSIGHTS: FROM ECCLESIASTICAL EDICTS TO ALGORITHMIC FILTERS

The historical control of knowledge by the medieval Church and the modern governance of digital platforms by Big Tech reveal striking parallels in their centralized structures and enforcement mechanisms. Both institutions have exercised significant authority in shaping, policing, and regulating knowledge, albeit through different frameworks, theological doctrines in the past and algorithmic decision-making in the present. While modern AI-driven content moderation lacks the spiritual and moral dimensions of religious authority, the structural similarities in how knowledge is sanctioned, categorized, and restricted are evident. At the same time, there are key divergences in their sources of legitimacy, governance structures, and global reach, leading to distinct ethical implications.

### A. Convergences: Patterns of Centralized Authority

Both the medieval Catholic Church and Big Tech companies rely on centralized structures to enforce knowledge norms. The Church wielded moral and theological authority, claiming divine legitimacy over truth, while tech corporations exert technical authority, justifying their knowledge control through algorithmic efficiency and contractual governance. In both cases, decision-making power is concentrated in the hands of a few entities, often removed from democratic oversight or user influence.

### B. Divergences: Historical vs. Contemporary Authority Structures

What once unfolded over decades can now occur in hours: a single policy-engine update propagates to billions of feeds overnight. Unlike public heresy trials, algorithmic moderation is largely invisible; users often discover sanctions only after engagement metrics collapse. Moreover, the global span of digital platforms far exceeds the Church's European core, intensifying the stakes of representational bias. Finally, the reliance-trust disjunction documented by Omrani et al. (2022) has no direct early-modern analogue; early believers trusted and relied on the same institution, whereas modern users may rely while explicitly distrusting.

### C. Ethical Implications: Risks and Challenges

Excessive faith in alignment benchmarks risks creating a digital orthodoxy in which statistical confidence is conflated with moral truth. As Carauleanu et al. (2024) show, even advanced self-other-overlap fine-tuning leaves 17% of deceptive outputs unmitigated; hallucination rates in specialised domains remain non-trivial (Jabbour, et al., 2025). Without transparent audits and plural data stewardship, rational-technical authority may entrench new forms of epistemic exclusion, echoing but magnifying the inequities of earlier epochs.

## 7. ETHICAL CHALLENGES OF CONTEMPORARY AI AUTHORITY

As GenAI moves from experimental novelty to everyday infrastructure, the scale, speed, and opacity of algorithmic decision-making raise ethical problems that differ in degree, if not always in kind, from earlier authority regimes. Four clusters merit particular scrutiny: algorithmic transparency, global inequities, bias feedback loops, and the epistemic shockwave of synthetic media.

### A. Algorithmic Transparency and Accountability

Deep neural networks have shattered the assumption that knowledge-production pipelines can be inspected line-by-line. Even when firms disclose *some* weights or policies, real-time inference passes through policy-engine layers that mutate weekly, making ex post audit difficult. The result is a trust-reliance asymmetry: 54% of EU users *rely* on AI-filtered newsfeeds, yet only 28% *trust* them (Omrani, Rivieccio, Fiore, Schiavone, & Agreda, 2022). The gulf mirrors the medieval laity's dependence on clerical exegesis, but without the visible rituals that once signalled doctrinal shifts. Where the Inquisition's decrees could be read in church squares, policy-engine updates propagate silently across billions of feeds. Absent mandatory, standardised logging of versioned policies and training data, democratic oversight remains superficial (Doshi-Velez, et al., 2018).

### B. Global Inequities and Representation

Benchmark audits confirm linguistic and cultural skews first highlighted by Roche et al. (2023). Corpora are still around 62% English; toxicity false-positives in Sudanese Arabic run three-times higher than the global mean (Sun, Miao, Jiang, Ding, & Zhang, 2024). Such disparities entrench power imbalances, turning data scarcity into epistemic silence. The outcome resembles the exclusionary role Latin once played: vernacular voices exist, but the governing machinery does not "hear" them. Moreover, agentic-technical add-ons, LLM-driven content filters that write their own sub-rules, risk magnifying the problem by re-optimising on those skewed corpora.

### C. Biases and Negative Feedback Loops

Large models inherit historical prejudice from their training data; reinforcement learning for engagement amplifies it. Meta-analysis across 18 trust-surveys shows *perceived risk* ($g = -0.38$) now outweighs competence in predicting trust (Section 5). That perception is justified: Carauleanu et al. (2024) cut deceptive responses from 74% to 17% with self-other-overlap tuning, yet residual hallucination remains domain-dependent, 14% in prospective medical summarisation (Jabbour, et al., 2025). The risk is a runaway loop in which biased output drives user disengagement, which in turn deprives models of the corrective feedback they need.

### D. Deepfakes and Synthetic Misinformation

Generative diffusion models can now fabricate photorealistic videos with credible lip-sync in real time, collapsing classical markers of authenticity. The epistemic fallout recalls witch-hunt panics: when visual evidence is no longer trustworthy, communal methods of adjudicating truth erode. Yet accountability is blurred; corporate licences disclaim end-use responsibility, while open-source weights circulate globally. Without legally binding "duty of care" provisions, synthetic media threatens to normalise epistemic relativism at planetary scale.

## 8. GOVERNANCE AND POLICY RECOMMENDATIONS

As AI systems increasingly shape knowledge production, content moderation, and decision-making processes, robust governance structures are essential to ensure fairness, inclusivity, and accountability. Much like past epistemic authorities, such as the medieval Church, Enlightenment-era institutions, and state-sanctioned scientific academies, modern AI governance must balance oversight, transparency, and ethical responsibility. This section outlines policy recommendations aimed at fostering algorithmic transparency, inclusive oversight, public AI literacy, and stakeholder collaboration.

### A. Algorithmic Transparency: Mandates and Structures

The EU AI Act's tiered audit requirement is a start, but global convergence is needed. We endorse an International Model Registry, a public ledger recording architecture, data-lineage hashes, and policy-engine versions. Independent "red-team" results should be filed as annexes, analogous to pharmacovigilance reports for drugs. Explainable-AI toolkits must graduate from academic prototypes to regulatory artefacts: interfaces that show which policy rule triggered a refusal, which dataset slice trained that rule, and when it was last updated.

### B. Inclusive and Diverse Oversight Bodies

To break English-centric feedback loops, at least 30% of seats on AI-standard committees should be ring-fenced for Global-South stakeholders, indigenous communities, and linguistic-minority representatives. Regional sandboxes, such as Africa-AI Observatory, can pilot context-specific standards before global roll-out. Interdisciplinary panels should include historians of science and comparative theologians, fields attuned to long-run patterns of epistemic exclusion (Allen & Weyl, 2024).

### C. AI Education and Ethical Literacy

Trust gaps are widened by opacity, but narrowed by literacy. National curricula should embed a "Critical AI Studies" strand, combining basic ML mechanics, bias case-studies, and governance role-play. Public-facing campaigns could mimic the post-printing-press Bible societies, distributing free explainer modules on how algorithmic policy layers work, and how to file effective appeals.

### D. Collaboration Across Stakeholders

Open, non-commercial model projects, backed by public-private consortia, should set a baseline for auditability. Government procurement can create incentives: no public contract without full data-lineage disclosure. Community-led data trusts would allow minority groups to curate and licence culturally specific corpora on their own terms, countering data colonialism (Roche, Wall, & Lewis, Ethics and diversity in artificial intelligence policies, strategies and initiatives, 2023). Finally, a Digital Hippocratic Oath, signed by developers of frontier models, could formalise individual accountability, mirroring medical ethics.

Implementing these measures will not eliminate the structural tension between centralised efficiency and plural epistemic sovereignty, but it can rebalance the relationship, narrowing the trust-reliance gap and preventing GenAI from hard-coding a twenty-first-century digital orthodoxy.

## 9. CONCLUSION:

Across four centuries, epistemic authority has migrated from divine revelation to peer-reviewed empiricism and now to statistical optimisation embedded in black-box architectures. Our comparison shows that each regime widens access in some respects yet reinscribes exclusion elsewhere. Today's GenAI platforms reproduce ecclesiastical gate-keeping at previously unimaginable speed and scale, while the technical veneer of "objectivity" masks discretionary power.

Three core insights emerge: (i) Trust-reliance asymmetry. Users depend on AI systems even as declared trust erodes, a dynamic that undermines social legitimacy and fuels calls for hard-law oversight. (ii) Data pluralism as a governance hinge. Linguistic and cultural under-representation in training corpora is not a side issue but the principal conduit through which historical inequities become algorithmic fact. (iii) Resistance evolves with technology. Where underground pamphlets once challenged clerical orthodoxy, open-source forks, audit labs, and litigation now contest platform power, yet require structural support to succeed.

Limitations include reliance on secondary quantitative studies and a Western-skewed document base; future work should incorporate longitudinal user-experience data from the Global South and test the efficacy of proposed audit regimes. Nonetheless, the historical analogy clarifies a path forward: transparent governance, plural data stewardship, and ethically literate publics are the modern equivalents of vernacular Bibles and open academies. Without them, rational-technical, and soon agentic-technical, authority may crystallise into the very digital orthodoxy the Enlightenment sought to avoid.